\documentclass[aps,paper]{revtex4}%
\usepackage{amsfonts}
\usepackage{amsmath}
\usepackage{amssymb}
\usepackage{graphicx}%
\begin{document}
\title[Regularization of the spectral problem for the monolayer graphene]{Regularization of the spectral problem for the monolayer graphene with the
separable in the angular momentum representation singular potential of defect}
\author{Sergey A. Ktitorov, Yurii I. Kuzmin,}
\affiliation{Ioffe Physical-Technical Institute, the Russian Academy of Sciences, St.
Petersburg, 194021, Russia}
\author{Natalie E. Firsova,}
\affiliation{Institute for Problems of Mechanical Engineering, the Russian Academy of
Sciences, St. Petersburg}
\keywords{2-spinor, separable potential}

\begin{abstract}
The electronic states in the monolayer graphene with the short-range
perturbation asymmetric with respect to the band index are analized. The study
was made for the separable in the angular momentum representation potential
basing on the (2+1)-dimensional Dirac equation. The characteristic equation
for bound and resonance states obtained in the present paper is compared with
one derived earlier for the same problem with different approach. The momentum
representation approach used in the present paper allowed us to obtain the
satisfactory regularization of the Hadamar incorrect boundary problem stemming
from the potential singularity.

\end{abstract}
\maketitle

%\pacs{81.05.Uw 72.10.-d 73.63.-b 73.40.-c}

E-mail: ktitorov@mail.ioffe.ru, nef2@mail.ru

%\volumeyear{year}
%\volumenumber{number}
%\issuenumber{number}
%\eid{identifier}
%\date[Date text]{date}
%\received[Received text]{date}
%\revised[Revised text]{date}
%\accepted[Accepted text]{date}
%\published[Published text]{date}
%\startpage{101}
%\endpage{102}
%\tableofcontents

\section{Introduction}

Transport theory in the presence of the resonances and theory of the optical
absorption in graphene need nonperturbative analysis of electronic states.
This makes it necessary to consider exactly solvable models of defects. One of
them is the delta function potential. In the case of the two-band
nonrelativistic problem for three-dimensional zero- and narrow-gap
semiconductors described by the Dirac equation the problem of bound and
resonance states was considered with a use of such potential in the paper
\cite{tamarshortrange}. The delta function is determined in this model on the
circumference of the sphere of proper dimension. In the present case it is the
circumference of circle. This potential has no singularity at $r=0$ and is
separable in the angular momentum representation. This potential is sepable in
the angular momentum representation. We take into account possible difference
of the perturbation matrix elements calculated on wave functions of the upper
and lower bands that is equivalent to consideration of both potential and mass
perturbations. Such asymmetry with respect to the band index can be induced by
the local shift of sublattices in the vicinity of the point defect. The aim of
the present work is a regularization of this incorrect in Hadamar's sence problem.

\section{Basic equations}

During the last years much attention was paid to the problem of the electronic
spectrum of graphene (see a review \cite{novosel}). Two-dimensional structure
of it and a presence of the cone points in the electronic spectrum make actual
a comprehensive study of the external fields effect on the spectrum and other
characteristics of the electronic states described by the Dirac equation in
the 2+1 space-time. Short-range potential impurities in graphene were
considered in works \cite{basko}, \cite{novikov}, \cite{matulis}. In our works
\cite{we1}, \cite{we}, a new model of the short-range impurities in graphene
was considered taking into account possible local shift of sublattices. This
means that the perturbation must be generically described by a Hermitian
matrix. We considered it in the diagonal representation. We do not take into
account the inter-valley transitions. The characteristic equation for the
bound states wasderived in \cite{we1} within the framework of this model. It
was understood that singularity of the delta-potential induces a problem of
incorrectness of the boundary problem that made some regularization necessary.
The present work is just dedicated to one possible way of regularization of
this Hadamar incorrect problem. Notwithstanding a presence of this difficulty,
the delta potential is extremely popular particularly for use in electronic
kinetics and, therefore, deserves thorough analysis.

The Dirac equation describing electronic states in graphene reads
\begin{equation}
\left(  -i\hbar v_{F}\sum_{\mu=1}^{2}\gamma_{\mu}\partial_{\mu}-\gamma
_{0}\left(  m+\delta m\right)  v_{F}^{2}{}\right)  \psi=\left(  E-V\right)
\psi, \label{diracgeneral}%
\end{equation}
where $v_{F}$ is the Fermi velocity of the band electrons, $\gamma_{\mu}$ are
the Dirac matrices%
\begin{equation}
\gamma_{0}=\sigma_{3},\text{ }\gamma_{1}=\sigma_{1},\text{ }\gamma_{2}%
=\sigma_{2}, \label{matrices}%
\end{equation}
$\sigma_{i}$ are the Pauli matrices, $2mv_{F}{}^{2}=E_{g}$ is the electronic
bandgap, $\psi\left(  \mathbf{r}\right)  $ is the two-component spinor:%

\begin{equation}
\psi\left(  \mathbf{r}\right)  =\left(
\begin{array}
[c]{c}%
f\left(  \mathbf{r}\right) \\
g\left(  \mathbf{r}\right)
\end{array}
\right)  . \label{spinorR}%
\end{equation}

. The electronic gap can appear in the graphene monatomic film lying on the
substrate because of the sublattices mutual shift \cite{gap}. The spinor
structure takes into account the two-sublattice structure of
graphene.$\ \delta m\left(  \mathbf{r}\right)  $ and $V(\mathbf{r})$ are the
local perturbations of the mass (gap) and the chemical potential. A local mass
perturbation can be induced by defects in a graphene film or in the substrate
\cite{gap}. We consider here the delta function model of the perturbation:%

\begin{equation}
\delta m\left(  \mathbf{r}\right)  =-b\delta(r-r_{0}),\text{ }V(\mathbf{r)}%
=-a\delta(r-r_{0}),\label{perturb}%
\end{equation}

that can be re-written in the form of the band-asymmetric potential%

\begin{equation}
V_{1}(\mathbf{r)=}V_{1}^{0}\delta(r-r_{0}),\text{ }V_{2}(\mathbf{r)=}V_{2}%
^{0}\delta(r-r_{0}).\label{pot}%
\end{equation}

The parameters $V_{1}^{0},$ $V_{2}^{0}$ , $a$ and $b$ are related as follows%

\begin{equation}
V_{1}^{0}=-\left(  a+b\right)  ,\text{ }V_{2}^{0}=-\left(  a-b\right)  .
\label{abVrelation}%
\end{equation}
Here $r$ and $r_{0}$ are respectively the polar coordinate radius and the
perturbation radius. Such short-range perturbation was used in the (3+1)-Dirac
problem for narrow-gap and zero-gap semiconductors in \cite{tamarshortrange}.
The perturbation forms the diagonal matrix
\begin{equation}
diag(V_{1}^{0},V_{2}^{0}) \label{diag}%
\end{equation}

The delta function perturbation is the simplest solvable short-range model.
Finite radius $r_{0}$ plays a role of the regulator and is necessary in order
to exclude deep states of the atomic energy scale. The finite perturbation
radius $r_{0}$ leads to the quasi-momentum space form-factor proportional to
the Bessel function that justifies our neglect of transitions between the
Brillouin band points $K$\ and $K^{\prime}$ \cite{tamarshortrange}.

Let us introduce the two-dimensional Fourier representation of the
two-component wave function (\ref{spinorR}:%

\begin{equation}
f\left(  \mathbf{r}\right)  =\int\frac{dp_{x}dp_{y}}{\left(  2\pi\right)
^{2}}e^{i\mathbf{kr}}f_{\mathbf{p}},\text{ \ }g\left(  \mathbf{r}\right)
=\int\frac{d^{2}p}{\left(  2\pi\right)  ^{2}}e^{i\mathbf{kr}}g_{\mathbf{p}},
\label{directfourier}%
\end{equation}

with the Fourier-components $f_{\mathbf{p}}$ and $g_{\mathbf{p}}$ determined
by the inverse Fourier transforms:%
\begin{equation}
f_{\mathbf{p}}=\int dxdye^{-i\mathbf{kr}}f\left(  \mathbf{r}\right)  ,\text{
\ }g_{\mathbf{p}}=\int dxdye^{-i\mathbf{kr}}g\left(  \mathbf{r}\right)
\label{inversefourier}%
\end{equation}

The inverse transform can be written as a combination of the Hankel transform
and the angular Fourier series \cite{morse}:%
\begin{equation}
f_{\mathbf{p}}\equiv f\left(  p,\theta\right)  =\sum_{n=-\infty}^{\infty}%
i^{n}e^{in\theta}f_{n}\left(  p\right)  ,\text{ \ \ }g_{\mathbf{p}}\equiv
g\left(  p,\theta\right)  =\sum_{n=-\infty}^{\infty}i^{n}e^{in\theta}%
g_{n}\left(  p\right)  \label{2dfourier}%
\end{equation}

\begin{equation}
f_{n}\left(  p\right)  =\int_{0}^{\infty}drrf_{n}\left(  r\right)
J_{n}\left(  pr\right)  ,\text{ \ \ }g_{n}\left(  p\right)  =\int_{0}^{\infty
}drrg_{n}\left(  r\right)  J_{n}\left(  pr\right)  \label{hankel}%
\end{equation}

where $f_{n}\left(  r\right)  $ is the angular Fourier component:
\begin{equation}
f\left(  r,\theta\right)  =\frac{1}{\sqrt{2\pi}}\sum_{n=-\infty}^{\infty}%
f_{n}\left(  r\right)  e^{in\theta},\text{ \ \ }g\left(  r,\theta\right)
=\frac{1}{\sqrt{2\pi}}\sum_{n=-\infty}^{\infty}g_{n}\left(  r\right)
e^{in\theta}\label{angular}%
\end{equation}

Making the Fourier transform of the Dirac equation (\ref{diracgeneral}), we
consider preliminary two terms of it, which can be considered as rather
complicated. The kinetic term reads%
\begin{equation}
-i\left(  \widehat{\sigma}_{x}\partial_{x}+\widehat{\sigma}_{y}\partial
_{y}\right)  \psi\left(  \mathbf{r}\right)  =\left(
\begin{array}
[c]{c}%
\left(  \widehat{p}_{x}+i\widehat{p}_{y}\right)  g\left(  \mathbf{r}\right) \\
\left(  \widehat{p}_{x}-i\widehat{p}_{y}\right)  f\left(  \mathbf{r}\right)
\end{array}
\right)  . \label{kinetic}%
\end{equation}

Therefore, we can write for the upper and lower components of the kinetic term
in the momentum representation
\begin{equation}
\left(
\begin{array}
[c]{c}%
\left(  p_{x}+ip_{y}\right)  g\left(  \mathbf{p}\right) \\
\left(  p_{x}-ip_{y}\right)  f\left(  \mathbf{p}\right)
\end{array}
\right)  =\left(
\begin{array}
[c]{c}%
pe^{i\theta}g\left(  p,\theta\right) \\
pe^{-i\theta}f\left(  p,\theta\right)
\end{array}
\right)  \label{kinetic2}%
\end{equation}

Substituting (\ref{2dfourier}) into (\ref{kinetic2}), we obtain:%
\begin{equation}
\left(
\begin{array}
[c]{c}%
+i\sum_{n=-\infty}^{\infty}i^{n}e^{in\theta}g_{n+1}\left(  p\right) \\
-i\sum_{n=-\infty}^{\infty}i^{n}e^{in\theta}f_{n-1}\left(  p\right)
\end{array}
\right)  \label{kinetic3}%
\end{equation}

The potential Fourier transform $V_{i}\left(  \mathbf{p}\right)  $ can be
expanded into a series in the case of the circular symmetry:%
\begin{equation}
V_{i}\left(  \left\vert \mathbf{p}-\mathbf{p}^{\prime}\right\vert \right)
=\pi\sum_{n=-\infty}^{\infty}e^{in\theta}\cdot V_{i}^{n}\left(  p,p^{\prime
}\right)  , \label{expansion}%
\end{equation}

\begin{equation}
V_{i}^{n}\left(  p,p^{\prime}\right)  =\int_{0}^{\infty}dr\cdot rJ_{n}\left(
pr\right)  V_{i}\left(  r\right)  J_{n}\left(  p^{\prime}r\right)
\label{expansion2}%
\end{equation}
where $J_{n}\left(  pr\right)  $ is the Bessel function, $i=1,$ \ \ $2$. Thus
we obtain the integral equations:

\begin{equation}
\left(  E-m\right)  f_{j}\left(  p\right)  +ipg_{j}\left(  p\right)  -\int
dp^{\prime}\cdot p^{\prime}f_{j}\left(  p^{\prime}\right)  V_{1}^{j}\left(
p^{\prime},p\right)  =0, \label{integral1}%
\end{equation}

\begin{equation}
\left(  E+m\right)  g_{j}\left(  p\right)  -ipf_{j}\left(  p\right)  -\int
dp^{\prime}\cdot p^{\prime}g_{j}\left(  p^{\prime}\right)  V_{2}^{j}\left(
p^{\prime},p\right)  =0, \label{integral2}%
\end{equation}

where we have put $n=j-1/2,$ where $j$ is the pseudospin quantumnumber,
$j=\pm1/2,$ $\pm3/2\ldots.$ In opposite to the relativistic theory, this
quantum number has nothing to do with the real spin and indicates a degeneracy
in the biconic Dirac point. These equations have a symmetry:
\begin{equation}
f_{j}\leftrightarrow g_{j},\text{ }E\rightarrow-E,\text{ }j-1/2\rightarrow
j+1/2,\text{ }a\rightarrow-a. \label{symm}%
\end{equation}
We have introduced the notations:%
\begin{equation}
f_{n=j-1/2}\equiv f_{j},\text{ \ }g_{n=j+1/2}\equiv g_{j}\text{\ }
\label{notation}%
\end{equation}

Zero-radius \cite{ostrov} and separable potentials \cite{newton} are popular
in the nonrelativistic scattering theory.\ However, the Dirac equation is
extremely sensitive to a singularity of the potential \cite{zeld}. The
singularity of the delta function potential can be regulated putting the delta
function to the circular support (\ref{perturb}) \cite{tamarshortrange}

Substituting (\ref{diag}) into (\ref{expansion}) we obtain the separable in
the angular momentum-momentum modulus representation potential:%
\begin{equation}
V_{i}^{j}\left(  p,p^{\prime}\right)  =v_{i}^{j}\left(  p\right)  v_{i}%
^{j}\left(  p^{\prime}\right)  , \label{separable}%
\end{equation}
where%

\begin{equation}
v_{1}^{j}\left(  p\right)  =\sqrt{r_{0}V_{1}^{0}}J_{j-1/2}(pr_{0}),\text{
\ \ }v_{2}^{j}\left(  p\right)  =\sqrt{r_{0}V_{2}^{0}}J_{j+1/2}(pr_{0}),i=1, 2
\label{factor}%
\end{equation}

.

Equations (\ref{integral1}), (\ref{integral2}) become degenerate and can be
written as follows:%

\begin{align}
\left(  E-m\right)  f_{j}\left(  p\right)  +ipg_{j}\left(  p\right)
-v_{1}^{j}\left(  p\right)  \int_{0}^{\infty}dp^{\prime}p^{\prime}f_{j}\left(
p^{\prime}\right)  v_{1}^{j}\left(  p^{\prime}\right)   &
=0,\label{separableeq1}\\
\left(  E+m\right)  g_{j}\left(  p\right)  -ipf_{j}\left(  p\right)
-v_{2}^{j}\left(  p\right)  \int_{0}^{\infty}dp^{\prime}p^{\prime}g_{j}\left(
p^{\prime}\right)  v_{2}^{j}\left(  p^{\prime}\right)   &  =0.
\label{separableeq2}%
\end{align}

\section{Characteristic equation}

Introducing the functions%
\begin{equation}
F_{j}\left(  E\right)  =\int_{0}^{\infty}dppf_{j}\left(  p\right)  v_{1}%
^{j}\left(  p\right)  ,\text{ \ \ }G_{j}\left(  E\right)  =\int_{0}^{\infty
}dppg_{j}\left(  p\right)  v_{2}^{j}\left(  p\right)  ,\text{ \ \ }R\left(
p\right)  =\left(  p^{2}+m^{2}-E^{2}\right)  ^{-1}, \label{FG}%
\end{equation}

we obtain the homogeneous algebraic equations set for $F$ and $G$:
\begin{align}
F_{j}\left(  1+\left(  E+m\right)  \int_{0}^{\infty}dppR\left(  p\right)
\left(  v_{1}^{j}\left(  p\right)  \right)  ^{2}\right)  -iG_{j}\int
_{0}^{\infty}dpp^{2}R\left(  p\right)  v_{1}^{j}\left(  p\right)  v_{2}%
^{j}\left(  p\right)   &  =0,\label{eqset1}\\
iF_{j}\int_{0}^{\infty}dpp^{2}R\left(  p\right)  v_{1}^{j}\left(  p\right)
v_{2}^{j}\left(  p\right)  +G_{j}\left(  1+\left(  E-m\right)  \int
_{0}^{\infty}dppR\left(  p\right)  \left(  v_{2}^{j}\left(  p\right)  \right)
^{2}\right)   &  =0. \label{eqset2}%
\end{align}

The solvability condition for this equations set gives the characteristic
equation:
\begin{align}
&  \left[  1+\left(  m+E\right)  \int_{0}^{\infty}dppR\left(  p\right)
\left(  v_{1}^{j}\left(  p\right)  \right)  ^{2}\right]  \cdot\left[
1+\left(  m-E\right)  \int_{0}^{\infty}dppR\left(  p\right)  \left(  v_{2}%
^{j}\left(  p\right)  \right)  ^{2}\right] \nonumber\\
&  =\left[  \int_{0}^{\infty}dpp^{2}R\left(  p\right)  v_{1}^{j}\left(
p\right)  v_{2}^{j}\left(  p\right)  \right]  ^{2}. \label{det}%
\end{align}

\bigskip Using the well known formula \cite{Prud}%
\[
\int_{0}^{\infty}dx\frac{x^{\mu-\nu+2n+1}}{x^{2}+z^{2}}J_{\mu}\left(
bx\right)  J_{\nu}\left(  cx\right)  =\left(  -1\right)  ^{n}z^{\mu-\nu
+2n}I_{\nu}\left(  x\right)  K_{\mu}\left(  x\right)
\]

we write it in the form:%

\begin{align}
&  \left[  1+\left(  m+E\right)  r_{0}V_{1}^{0}I_{j-1/2}\left(  \kappa
r_{0}\right)  K_{j-1/2}\left(  \kappa r_{0}\right)  \right]  \left[  1+\left(
m-E\right)  V_{2}^{0}I_{j+1/2}\left(  \kappa r_{0}\right)  K_{j+1/2}\left(
\kappa r_{0}\right)  \right] \nonumber\\
=\left(  m+E\right)  \left(  m-E\right)   &  V_{1}^{0}V_{2}^{0}r_{0}%
^{2}I_{j-1/2}^{2}\left(  \kappa r_{0}\right)  K_{j+1/2}^{2}\left(  \kappa
r_{0}\right)  , \label{character}%
\end{align}

where $\kappa^{2}=\left(  m^{2}-E^{2}\right)  ,$ $I_{n}\left(  x\right)  ,$
$K_{n}\left(  x\right)  $ are the modified Bessel functions. Making use of the
identity \cite{grad}

\begin{equation}
I_{\nu}\left(  x\right)  K_{\nu+1}\left(  x\right)  +I_{\nu+1}\left(
x\right)  K_{\nu}\left(  x\right)  =1/x, \label{identity}%
\end{equation}

we obtain the characteristic equation:

\begin{align}
&  \kappa\left[  I_{l-1/2}\left(  \kappa r_{0}\right)  K_{j+1/2}\left(  \kappa
r_{0}\right)  +I_{j+1/2}\left(  \kappa r_{0}\right)  K_{j-1/2}\left(  \kappa
r_{0}\right)  \right]  +\left(  m+E\right)  V_{1}^{0}I_{j-1/2}\left(  \kappa
r_{0}\right)  K_{j-1/2}\left(  \kappa r_{0}\right)  +\nonumber\\
&  \left(  m-E\right)  V_{2}^{0}I_{j+1/2}\left(  \kappa r_{0}\right)
K_{j+1/2}\left(  \kappa r_{0}\right)  =0. \label{charfinal1}%
\end{align}

Using the relations (\ref{abVrelation}), we can write the characteristic
equation in the form:

\[
\kappa\left[  I_{j-1/2}\left(  \kappa r_{0}\right)  K_{j+1/2}\left(  \kappa
r_{0}\right)  +K_{j-1/2}\left(  \kappa r_{0}\right)  I_{j+1/2}\left(  \kappa
r_{0}\right)  \right]  =
\]

\begin{equation}
\left[  (m-E)\left(  a-b\right)  I_{j+1/2}\left(  \kappa r_{0}\right)
K_{j+1/2}\left(  \kappa r_{0}\right)  +\left(  a+b\right)  (m+E)I_{j-1/2}%
\left(  \kappa r_{0}\right)  K_{j-1/2}\left(  \kappa r_{0}\right)  \right]  .
\label{character1a}%
\end{equation}

This equation has to be compared with one derived using the different approach
in \cite{we}:

\begin{align}
&  p\left(  J_{j-1/2}\left(  pr_{0}\right)  H_{j+1/2}^{\left(  1\right)
}\left(  pr_{0}\right)  -J_{j+1/2}\left(  pr_{0}\right)  H_{j-1/2}^{\left(
1\right)  }\left(  pr_{0}\right)  \right) \nonumber\\
&  =\frac{\tan\left(  \sqrt{a^{2}-b^{2}}\right)  }{\sqrt{a^{2}-b^{2}}}\left[
\sqrt{\frac{E-m}{E+m}}\left(  a-b\right)  J_{j+1/2}\left(  pr_{0}\right)
H_{j+1/2}^{\left(  1\right)  }\left(  pr_{0}\right)  +\sqrt{\frac{E+m}{E-m}%
}\left(  a+b\right)  J_{j-1/2}\left(  pr_{0}\right)  H_{j-1/2}^{\left(
1\right)  }\left(  r_{0}\right)  \right]  , \label{char2}%
\end{align}
where $H_{n}^{\left(  1\right)  }\left(  z\right)  $ is Hankel's function,
$p=\sqrt{E^{2}-m^{2}}.$ Making the analytic continuation from the case of the
band states with $E^{2}>m^{2}$ to the opposite one of bound states
$E^{2}<m^{2}$ we obtain the equation%

\[
\kappa\left[  I_{j-1/2}\left(  \kappa r_{0}\right)  K_{j+1/2}\left(  \kappa
r_{0}\right)  +K_{j-1/2}\left(  \kappa r_{0}\right)  I_{j+1/2}\left(  \kappa
r_{0}\right)  \right]  =
\]

\begin{equation}
\frac{\tan\left(  \sqrt{a^{2}-b^{2}}\right)  }{\sqrt{a^{2}-b^{2}}}\left[
(m-E)\left(  a-b\right)  I_{j+1/2}\left(  \kappa r_{0}\right)  K_{j+1/2}%
\left(  \kappa r_{0}\right)  +\left(  a+b\right)  (m+E)I_{j-1/2}\left(  \kappa
r_{0}\right)  K_{j-1/2}\left(  \kappa r_{0}\right)  \right]  . \label{old}%
\end{equation}

We see that the only distinction between the formulae (\ref{character1a}) and
(\ref{old}) is a presence of the factor
\begin{equation}
T\left(  a,b\right)  =\frac{\tan\left(  \sqrt{a^{2}-b^{2}}\right)  }%
{\sqrt{a^{2}-b^{2}}} \label{T}%
\end{equation}

in the equation derived in \cite{we}. They are identical in the limit
\begin{equation}
a^{2}-b^{2}\longrightarrow0,\text{ \ \ }T\left(  a,b\right)  \longrightarrow1.
\label{tan}%
\end{equation}
This limit can be reached near the lines $a^{2}-b^{2}=0$, or when $a$ and $b$
are small. The reason of this distinction is the following. The delta function
potential in the Dirac equation creates a boundary problem, which is incorrect
in the Hadamar sense \cite{payne}. Such a problem needs a regularization.
Distinct regularization procedures were used in the present work and in
\cite{we}. Singular feature of the potential was preserved in \cite{we}, a
regularization was carried out due to very special join of partial solutions,
while in the present work, a regularization took place due to the Fourier
transform smoothing property. In particular, the potential takes the separable
form with the nonlocal kernel $v_{i}^{j}\left(  p\right)  =\sqrt{r_{0}%
V_{i}^{0}}J_{j+1/2}(pr_{0})$ (see eq. (\ref{factor})). Just this nonlocality
plays the role of regularizator. Equations (\ref{character1a}) and (\ref{old})
have asymptotically similar solutions , when (\ref{tan}) is valid. Their sets
of solutions are quite different outside this region of parameters $a$ and
$b.$ Moreover, equation (\ref{old}) has significantly richer set of solutions,
than equation (\ref{character1a}) \cite{we} due to periodicity of tangent
present in (\ref{T}). We come to conclusion that the rich set of solutions
given by (\ref{old}) is the artifact of the zero-support interaction. Besides,
the annular well model analyzed in \cite{weftt} does not reproduce all of this
rich set of solutions even in the limit of zero radius. Therefore, the
equation obtained in the present work can be considered as a correctly
regularized one.

\section{Conclusion}

Non-relativistic problem of the electronic spectrum described by the Dirac
equation is considered in the case of the two-component short-range
perturbation potential with a help of the Fourier transforming. This approach
is shown to make regularized the incorrect in the Hadamar sense problem.


\begin{thebibliography}{99}                                                                                               %


\bibitem {tamarshortrange}V.I. Tamarchenko, S.A. Ktitorov, Soviet Physics --
Solid State, \textbf{19}, 2970 (1977).

\bibitem {novosel}A.H. Castro Neto, F. Guinea, et al, arXiv: 0709.1163 (2008).

\bibitem {basko}D.M. Basko, Phys. Rev. B \textbf{78}, 115432 (2008).

\bibitem {novikov}D.S. Novikov, Phys. Rev. B \textbf{76}, 245435 (2007).

\bibitem {matulis}A. Matulis, F.M. Peeters, Phys. Rev. B \textbf{77}, 115423 (2008).

\bibitem {we1}Natalie E. Firsova, Sergey A. Ktitorov, Philip A. Pogorelov,
Physics Letters A\textbf{ 373}, 525 (2009).

\bibitem {we}Natalie E. Firsova, Sergey A. Ktitorov, Physics Letters A
\textbf{374}, 1270 (2010).

\bibitem {gap}Aurelien Lherbier, X. Blaze, et al, Phys. Rev. Letters,
\textbf{101}, 036808-1 (2008).

\bibitem {morse}Philip M. Morse, Herman Feshbach, \textit{Methods of
Theoretical Physics, McGraw-Hill Book Company, }New York, Toronto, London, 1953.

\bibitem {ostrov}Yu.N. Demkov, V.N.Ostrovsky, \textit{Zero-radius Potentials
in Atomic Physics}, University Publishing House, Leningrad, 1975.

\bibitem {newton}Roger G. Newton, \textit{Scattering theory of waves and
particles}, Springer-Verlag, New York, 1982.

\bibitem {zeld}Ya.B. Zeldovich and V.S. Popov, Soviet Physics-Uspekhi,
\textbf{14}, 673 (1972).

\bibitem {Prud}A.P. Prudnikov, Yu.A. Brychkov, O.I. Marichev,
\textit{Integrals and series.Special functions}, Moscow, Nauka, 1983.

\bibitem {grad}I.S. Gradshtein, I.M. Ryzhik, \textit{Tables of integrals,
sums, series and products}, Fizmat Publishing, Moscow, 1963.

\bibitem {payne}L.E. Payne, \textit{Some General Remarks on Improperly Posed
Problems for Partial Differential equations.} In Lecture Notes in Mathematics
\textbf{316}, Symposium on Non-Well-Posed Problems and Logarithmic Convexity.
Edited by A. Dold and B. Eckmann, Edinburg, 1972.

\bibitem {weftt}Natalie E. Firsova, Sergey A. Ktitorov, Solid State Physics
(St. Petersburg) \textbf{53}, 376 (2011).
\end{thebibliography}
\end{document}